# Enhancing Cloud Security through Topic Modelling

Sabbir M. Saleh
Department of Computer Science
University of Western Ontario
London, ON, Canada
ssaleh47@uwo.ca
https://orcid.org/0000-0001-9944-2615

Nazim Madhavji
Department of Computer Science
University of Western Ontario
London, ON, Canada
madhavji@gmail.con
https://orcid.org/0009-0006-5207-3203

John Steinbacher
IBM Canada Lab.
Markham, ON, Canada
jstein@ca.ibm.com
https://orcid.org/0009-0001-6572-6326

*Abstract*—Protecting cloud applications is critical in an era where security threats are increasingly sophisticated and persistent. Continuous Integration and Continuous Deployment (CI/CD) pipelines are particularly vulnerable, making innovative security approaches essential. This research explores the application of Natural Language Processing (NLP) techniques, specifically Topic Modelling, to analyse security-related text data and anticipate potential threats. We focus on Latent Dirichlet Allocation (LDA) and Probabilistic Latent Semantic Analysis (PLSA) to extract meaningful patterns from data sources, including logs, reports, and deployment traces. Using the Gensim framework in Python, these methods categorise log entries into security-relevant topics (e.g., phishing, encryption failures). The identified topics are leveraged to highlight patterns indicative of security issues across CI/CD's continuous stages (build, test, deploy). This approach introduces a semantic layer that supports early vulnerability recognition and contextual understanding of runtime behaviours.

## I. INTRODUCTION

In today's rapidly changing digital world, the security of cloud-based applications is a crucial concern for organisations worldwide. With the growth of cloud services and the increasing complexity of software infrastructures, defence against cyber threats becomes increasingly tricky.

One critical area of focus inside this domain is the Continuous Integration and Continuous Deployment (CI/CD) pipeline. To ensure the security of cloud applications, vulnerabilities must be identified and addressed before deployment.

In the current state of software development, continuous integration relates to continuous deployment, which eliminates the discontinuity of development and deployment [1]. So, first comes to build, test, and last comes to deploy. A CI/CD pipeline is simplifying these tasks.

However, the CI/CD pipeline has some security issues (such as data privacy [2], exploited scripts [3], lack of authentication [4], etc.).

Additionally, ensuring security in cloud environments is a continual challenge. Protecting cloud platforms from various security breaches, such as downgrade attacks on Transport Layer Security (TLS), a crypto protocol (e.g., ROBOT, DROWN, POODLE) [5] and software supply chain attacks (e.g., Log4j, CodeCov, SolarWinds) [6, 7] is crucial. Failure to recognise and manage these risks can have serious consequences.

This study focuses on applying Topic Modelling techniques, such as Latent Dirichlet Allocation (LDA) and Probabilistic Latent Semantic Analysis (pLSA) of Natural Language Processing (NLP), to improve the overall security of cloud-based applications throughout the CI/CD pipeline. By utilising the power of natural language processing and machine learning, the goal is to develop a framework capable of predicting and exposing potential security vulnerabilities in text-based data generated throughout the CI/CD pipeline.

The primary outcome of this study is to develop a novel framework that utilises Topic Modelling, particularly LDA, and cosine similarity techniques to analyse text-based data such as plain text, lack of integration, disorganised contents, lack of contexts, partial incident reports, truncated logs, or isolated pieces of information from the CI/CD pipeline.

By comparing text-based documents against selected datasets containing security-related information, the framework identifies patterns that resemble security vulnerabilities, allowing users to identify and address potential risks before deployment.

Utilising synonyms (for example, protection, risk, threat, virus, spyware, etc.) and associated terminology (such as encryption, phishing, hacking, identity theft, etc.), we seek to uncover hidden and uncovered insights, classify the incidents, and strengthen security intelligence.

The fundamental question driving this research is:

RQ: How can Topic Modelling techniques be effectively applied to analyse fragmented security-related text data and enhance the security of cloud-based applications throughout the CI/CD pipeline?

This research brings a fresh approach by integrating Topic Modelling into the CI/CD pipeline to detect security threats. Its significance lies in connecting natural language processing techniques with cloud security, making it easier to share threat intelligence and identify risks throughout development.

The remainder of this paper is structured as follows:

Section II – Background and Related Works: Reviews existing research on cloud security and topic modelling, identifying key challenges and gaps addressed by this study.

Section III – Research Objectives: Defines the study's aims and outlines the specific goals pursued.

Section IV – Methodology: Describes the development steps and techniques used to construct the proposed framework.

Section V – Results: Presents the core findings and demonstrates how the framework enhances cloud security.

Section VI – Discussion: Analyses the results, addressing limitations, implications, and broader relevance.

Section VII – Conclusion: Summarises key insights, highlights contributions, and outlines directions for future work.

## II. RELATED WORKS

As security threats targeting CI/CD pipelines continue to evolve, researchers have explored various methods to extract meaningful insights from large volumes of development and deployment data. One such approach is topic modelling, a natural language processing (NLP) technique that uncovers latent themes in unstructured text. It has been widely applied in fields ranging from software engineering to cybersecurity, demonstrating its potential to interpret complex log data, identify threat patterns, and support proactive defence mechanisms. This section reviews prior work on topic modelling applications in security contexts, the development of supporting tools, and recent efforts to integrate semantic analysis and blockchain frameworks for enhancing runtime observability in DevSecOps environments.

In [8], Jelodar et al. (2019) explored applications of Topic Modelling, specifically LDA, across fields such as, medical science, software engineering, and political science. Reviewing papers from 2003 to 2016 demonstrated the versatility of topic modelling in extracting meaningful insights across various disciplines. Their work also shows its potential for analysing security-related text data to enhance cloud security.

Paradis et al. (2018) [9] used LDA to classify software vulnerability data from the Common Vulnerabilities and Exposures (CVE) project. Their approach required minimal preprocessing and helped organise complex, noisy data to extract valuable insights, providing a novel method for navigating large amounts of vulnerable information.

Okey et al. (2023) [10] analysed Twitter discussions on ChatGPT and cybersecurity, revealing mixed opinions about its potential benefits and risks. Using LDA, they categorised the discussions under hashtags such as, #chatgptsecurity, demonstrating how topic modelling can be applied to analyse historical security data.

Řehůřek et al. (2010) [11] introduced GENSIM, an NLP framework focused on scalability and ease of use for topic modelling. GENSIM supports Latent Semantic Analysis (LSA) and LDA, regardless of the training corpus size. This framework is essential for our research, offering a scalable and user-friendly platform for applying topic modelling in cloud security.

Roepke (2022) [12] and Ruchirawat (2020) [13] both highlight practical applications of Topic Modelling, emphasising data preparation techniques that enhance dataset coherence and relevance. Their insights support our approach to managing fragmented security-related data, forming a foundation for utilising topic modelling in cloud security enhancements.

A recent systematic literature review [31] highlights ongoing limitations in CI/CD pipeline security, including fragmented tool adoption, weak runtime visibility, and lack of semantic insight into pipeline activities. While traditional defences rely on static rules or audit controls [32], there is increasing demand for approaches that can extract context from pipeline behaviour logs and adapt to evolving threats.

In parallel, blockchain-based security frameworks have emerged as promising strategies to ensure tamper-evident, traceable logging [33] of pipeline activities. These approaches treat CI/CD operations as modular, verifiable blocks, helping isolate and contain threats before they escalate [30].

### A. Analysis

The reviewed literature provides various insights into using Topic Modelling in cybersecurity. Okey et al. (2023) illustrated the effectiveness of LDA in analysing ChatGPT discussions [10]. Paradis et al. (2018) used LDA to classify software vulnerability data, contributing to improved cybersecurity practices [9]. Jelodar et al. (2019) emphasised the value of Topic Modelling in extracting insights across diverse fields [8]. Furthermore, Řehůřek et al. (2010) developed GENSIM to tackle scalability challenges in NLP [11].

Recent studies have also expanded the security landscape in CI/CD contexts. A systematic literature review [32] identified key limitations in current CI/CD security practices, such as fragmented tooling, limited semantic observability, and a lack of runtime threat detection. Parallel to this, blockchain-based security frameworks [35] have been proposed to ensure tamper-evident logging and threat isolation in software pipelines.

However, a clear gap exists in applying topic modelling directly to cloud security within CI/CD pipelines. Most studies focus on analysing historical data, lacking proactive approaches to predict vulnerabilities in real-time environments. While GENSIM is scalable, its direct application in cloud security is still largely unexplored.

This study aims to bridge that gap by using Topic Modelling to identify security issues throughout the CI/CD pipeline proactively.

## III. RESEARCH OBJECTIVES

This research aims to investigate the potential of topic modelling techniques to enhance cloud security in CI/CD pipelines. The specific objectives are as follows:

O1: Explore the application of Topic Modelling techniques, particularly LDA, in analysing security-related text data.

O2: Investigate the effectiveness of Topic Modelling in predicting potential vulnerabilities in cloud applications.

O3: Provide insights and recommendations for integrating Topic Modelling into the CI/CD pipeline to improve overall cloud security throughout the software development process.

O4: Develop a scalable and user-friendly framework for applying Topic Modelling techniques to enhance cloud security practices throughout the CI/CD pipeline.

## IV. RESEARCH METHODOLOGY

In developing our solution, we adopted a hybrid approach that combines empirical research methodologies with lightweight software development practices. We detailed some of those in Section V-B.

The key to our approach was using topic modelling, specifically LDA and cosine similarity techniques, to analyse data and identify security-related patterns.

LDA is a generative probabilistic model used to discover latent topics within a collection of documents. It works by assuming that each document is a mixture of various topics, and each word is attributed to one of those topics [14].

Cosine similarity measures the similarity between two vectors by calculating the cosine of the angle between them [15]. Its simplicity and efficiency in measuring similarity between documents allow the system to compare documents and datasets effectively.

To validate our framework, we employed two datasets:

- The Worldwide Software Supply Chain Attacks Tracker (Comparitech) [16]
- Breaking Trust: Shades of Crisis across an Insecure Software Supply Chain (Herr, 2021) [17]

Both datasets provide curated web links to a range of cloud security-related sources, including technical blogs, whitepapers, breach reports, etc.

These real-world texts provide diverse, unstructured inputs that are ideal for evaluating topic modelling effectiveness in identifying security-relevant patterns across the CI/CD pipeline.

## V. RESULTS AND ANALYSIS

This section presents the practical implementation and evaluation of the proposed framework for enhancing cloud security within CI/CD pipelines. By integrating topic modelling and document similarity analysis into a scalable web-based platform, the system enables early detection of vulnerabilities in security-related text data, such as release notes and logs. The following subsections detail the system's architecture, functionality, integration into the CI/CD pipeline, and the results obtained through rigorous testing and validation procedures.

### A. CI/CD Pipeline Integration

The proposed system enhances cloud security practices by analysing text-based data generated throughout the CI/CD pipeline.

It consists of several core components, including a web application for user interaction, a backend server for data processing, a MySQL database for data storage, and algorithms for topic modelling and vulnerability detection.

Users can upload text-based files, e.g., release notes, log files, configuration summaries, etc. from the CI/CD pipeline for the vulnerability analysis. The system interfaces with the CI/CD pipeline before deployment, identifying potential security risks at critical stages of the software development lifecycle (Figure 1).

### B. System Design and Architecture

The system is built on a client-server architecture, which enhances modularity, scalability, and security. REST APIs facilitate seamless communication between the client and server, ensuring efficient data exchange and smooth operation. This design separates the user interface (client-side) from application logic and data storage (server-side), allowing independent updates and modifications to each component.

The architecture also supports resource distribution across multiple servers, enabling scalability to handle increased loads. Clustering sensitive data and logic on the server side reduces the risk of unauthorised access, further strengthening the system's security.

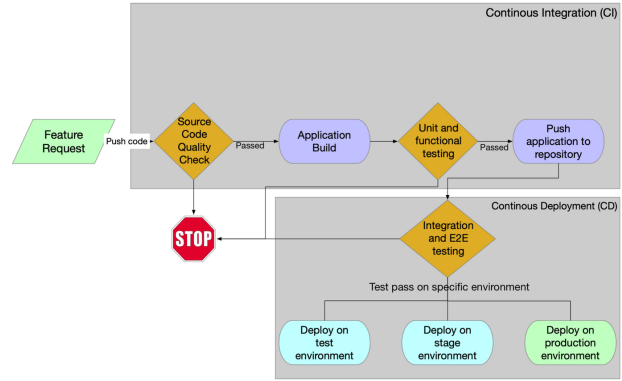

Fig. 1. Integration of the Proposed System within the CI/CD Pipeline

### C. Technical Implementation

- Web Application

The web application serves as the primary user interface, enabling users to:

a) Upload files or datasets for analysis.
b) Initiate the vulnerability detection process.
c) View results in an intuitive and user-friendly format.

- Backend Functionality

The backend server handles all data processing tasks, leveraging Python for its versatility and robust NLP support. Key backend components include:

*a) Topic Modelling:* LDA extracts latent topics from uploaded datasets, enabling the analysis of fragmented security-related text data.

*b) Vulnerability Detection:* The system compares CI/CD pipeline input files to uploaded datasets, identifying related security incidents. It checks the relevance of databases using topic modelling results and employs cosine similarity to find the top 10 most similar documents.

- Data Management

Data is stored in a MySQL database, chosen for its reliability, efficiency, and robust capabilities. Serialised LDA models are stored to optimise performance, allowing quick retrieval and reuse without repeated topic modelling.

*D. Key Features*

- Security

User passwords are encrypted to protect credentials and safeguard sensitive information.

Clustering sensitive data on the server minimises the risks of unauthorised access.

- Scalability and Performance

Serialised LDA models reduce processing overhead, improving responsiveness when handling large datasets.

The system architecture supports resource distribution, ensuring scalability as workloads increase.

*E. CI/CD Pipeline Integration and Results*

The system's integration into the CI/CD pipeline enables the detection of security vulnerabilities before the deployment stage. The system analyses text-based data and exposes potential risks in files such as release notes or logs.

This result validated the system's ability to detect vulnerabilities and share threat intelligence.

The following algorithm (Figure 2 presents the pseudocode) was designed to efficiently identify security vulnerabilities by analysing the similarity between the input file and existing datasets.

This system's development involved several libraries and tools, each chosen to fulfil a specific function and optimise performance. The tools range from backend frameworks such as, Flask and SQLAlchemy to data processing libraries such as Pandas and Gensim and frontend technologies such as, TypeScript and React. These components played a vital role in creating a robust, scalable, and efficient platform.

Table 1 lists these technologies, describing their purposes and benefits to the system.

TABLE I. LIBRARIES AND TOOLS USED

| Technology | Purpose | Features/Benefits |
|---|---|---|
| Flask [18] | Web framework for building RESTful APIs and handling HTTP requests efficiently. | Lightweight and flexible, suitable for RESTful API development. |
| SQLAlchemy [19] | ORM tool for database management. | High-level interface for interacting with the MySQL database. |
| Gensim [11] | Topic modelling tasks and text analysis. | Efficient implementations of LDA and other algorithms; scalable and user-friendly. |
| Pandas [20] | Data preprocessing and dataset formatting. | Versatile in data manipulation and preprocessing; enables efficient dataset handling. |
| Sci-kit-learn (sklearn) [21] | Machine learning tasks and cosine similarity calculations. | Robust ML functionalities ensure accurate document and dataset comparisons. |
| NLTK [22] | Text processing tasks such as tokenisation and stop word removal. | Enhances accuracy and effectiveness of text analysis. |
| TypeScript [23] | The primary language for frontend implementation. | Strong typing and scalability for easy codebase maintenance and development. |
| React [24] | Frontend framework for building user interfaces. | Component-based architecture offers a wide range of reusable UI components. |
| Material UI [25] | React component library for frontend design and styling. | Provides a modern and visually appealing user interface. |
| MySQL [26] | Database management system for data storage. | Reliable, scalable, and widely used; ensures data integrity and consistency. |
| Redis [27] | Session token storage for server-sided sessions. | Fast and scalable session management capabilities. |

```
Algorithm Topic_Modeling_and_Similarity_Analysis
Input: filename
Output: Topics, Tokenized Data, LDA Model

1. Load the Excel file (filename) containing References and
Metadata
2. For each row in filename:
    a. Retrieve webpage content from the URL.
    b. Clean and preprocess text:
        i. Remove punctuation, numbers, and stopwords.
        ii. Convert text to lowercase.
        iii. Apply lemmatization to retain nouns and adjectives.
    c. Store the cleaned text in the column "Summary".
3. Tokenize the cleaned text and store it in "Tokens".
4. Create a dictionary of tokens and a document-term matrix.
5. Apply LDA to the document-term matrix to discover latent topics:
    a. Define num_topics (e.g., 10).
    b. Train LDA with the tokens.
6. Save the trained LDA model and results to a file.

End
```

Fig. 2. Pseudocode for Topic Modelling and Similarity Analysis Algorithm

- System Testing

Multiple testing methodologies were utilised to ensure the system's reliability, functionality, and performance across various use cases. Table 2 lists those.

TABLE II. TESTING METHODOLOGIES AND THEIR PURPOSES

| Testing Methodology | Purpose | Key Features/Benefits |
|---|---|---|
| Unit Testing | Verifies that individual backend components operate correctly and meet specifications. | Ensures correctness and reliability of isolated functionalities. |
| Integration Testing | Ensures the system functions seamlessly when new features are integrated. | Validates interactions between different components of the system. |
| Acceptance Testing | Assesses the system against initial requirements and expectations. | Confirms that the final system meets user needs and design goals. |
| API Testing with Postman | Tests API endpoints to ensure correct responses to various inputs and scenarios. | Validates API functionality and identifies potential issues in communication layers. |

*F. System Validation*

The implemented system was validated to ensure its effectiveness and reliability in detecting security vulnerabilities throughout the CI/CD pipeline.

To test its performance, a case study was conducted using GitHub's latest release notes (Enterprise Server 3.10.9). A text document containing the release notes was uploaded to the system and compared against its datasets using the built-in algorithm.

The analysis identified websites associated with security attacks, with a similarity score exceeding 60%, demonstrating the system's capability to expose potential security risks in newly released software updates.

*G. System Walkthrough*

The system offers a streamlined, user-friendly interface designed to assist in identifying potential security vulnerabilities efficiently. Users can upload text-based files or datasets for analysis upon logging in or registering. The system allows users to define specific comparison parameters, such as selecting relevant datasets or setting similarity thresholds, to tailor the analysis to their needs. Once the analysis is complete, the system generates a detailed report highlighting potential vulnerabilities and providing links to related documents and websites. This intuitive process ensures that users can effectively uncover security risks in files and datasets, enhancing the security of CI/CD pipelines.

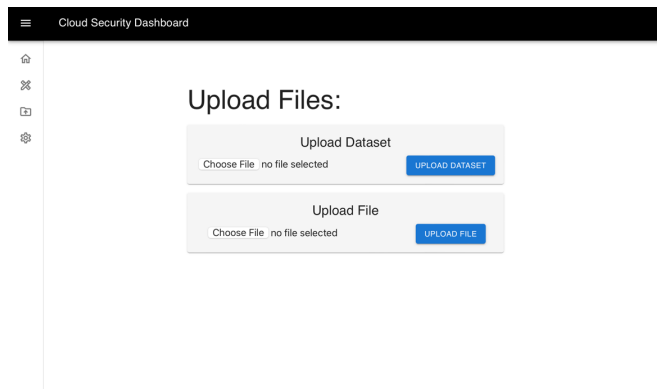

Fig. 3. File and Dataset Upload Interface in the Cloud Security System

The first step in using the system involves uploading files or datasets for analysis. Users can click the "Upload Files" button to select relevant CI/CD pipeline documents or choose the "Upload Datasets" button to upload entire datasets for comparison. As illustrated in Figure 3, this step allows users to provide input data efficiently for security analysis.

Users navigate to the Comparison page once the files or datasets are uploaded. Here, they select a file from the uploaded list and choose one or more datasets to compare against the selected file. Users initiate the analysis process by clicking the "Run Model" button. This step, shown in Figure 4, enables the system to analyse and compare the documents, facilitating the identification of potential security vulnerabilities.

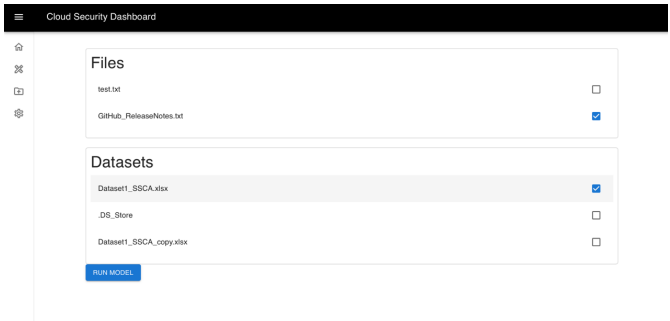

Fig. 4. File and Dataset Selection Prior to Model Execution

In the final step, users are directed to the Results page, where the system displays the top 10 most similarities from the selected datasets. Each result includes key details such as the dataset name, document link, and similarity percentage. These insights allow users to perform in-depth analysis and take appropriate actions based on the findings. This step is illustrated in Figure 5.

*H. Novelty of Results*

This study takes a fresh approach to improving cloud application security by focusing on analyzing text-based data within the CI/CD pipeline.

A vital feature of this method is the proactive use of topic modelling, which helps uncover security vulnerabilities before deployment.

The platform analyses fragmented security-related text data from multiple datasets through Topic Modelling, uncovering similarities and potential risks in documents for example, release notes and logs. This integration enables the platform to share actionable threat intelligence with users effectively.

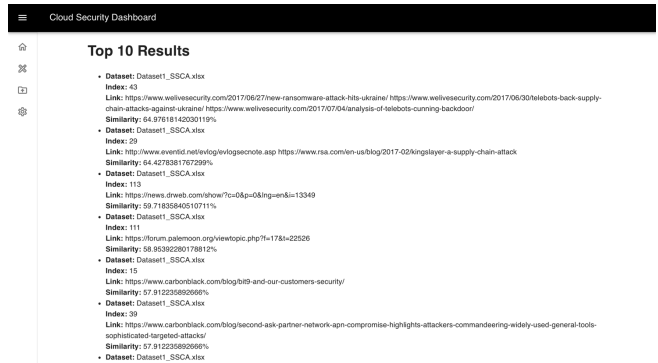

Fig. 5. Top 10 Matched Results with Similarity Scores

## VI. DISCUSSION

This section reflects on the practical implications, limitations, and broader impact of the proposed system. Building on the results, we evaluate the threats to validity, potential for generalisation, and pathways toward deeper integration into secure DevSecOps practices. The discussion also highlights how semantic insights from topic modelling can contribute to a multi-layered defence approach in modern CI/CD pipelines.

*A. Threats to the validity of results*

Several factors may affect the validity of the results, including the quality of datasets, which could suffer from broken links, outdated information, or irrelevant content. Uploaded files might also contain noise, such as unnecessary details or duplicates, skewing the analysis. Documents returned during file comparisons may sometimes fail to align with the intended research goals, which could lead to misleading conclusions.

To address these issues, preprocessing methods such as text cleaning, removing stop words, and lemmatization are applied to improve data quality by reducing noise. Despite these efforts, some risks remain, particularly regarding the relevance of the results. Regular manual review and ongoing algorithm refinements are essential to ensure the system's reliability.

Continuous enhancements will further improve the accuracy and consistency of the findings.

*B. Implication of the results*

This study contributes to both research and practice in cloud security by introducing a novel approach to detecting vulnerabilities through text-based data analysis in the CI/CD pipeline.

This practical approach helps link research insights to real-world applications, providing a solid foundation for enhancing cloud security practices and safeguarding applications against new and evolving threats.

Integrating this topic modelling approach with real-time anomaly detection systems [28] and blockchain-based logging mechanisms [30] may significantly improve the transparency and trustworthiness of CI/CD pipelines. This layered security design aligns with recent trends in secure DevSecOps, as highlighted by current literature reviews [31].

*C. Limitations of results*

While the developed system demonstrates significant potential for enhancing cloud security practices, certain limitations must be addressed to realise its capabilities fully:

Scope and Coverage:

The system's ability to perform effectively depends on having a broad and reliable range of security-related data sources. If the datasets are outdated or limited, important threats might be missed. To address this, it's crucial to expand the variety of data sources and keep the repository updated regularly, ensuring better detection and management of security risks.

Resource Limitations:

Another challenge is the impact of limited computational resources on scalability and performance, especially when handling large volumes of text data. These constraints can affect efficiency under heavy workloads. Enhancing the system's architecture and investing in advanced infrastructure, such as distributed computing, can help overcome these limitations and improve processing capabilities.

To minimise the impact of these limitations, the following measures are recommended:

Enhanced Data Coverage:

Broadening data coverage and regularly refreshing datasets can provide a more accurate and up-to-date analysis of security threats. These improvements are critical for ensuring effective risk detection and mitigation.

Resource Allocation:

Investing in better computational resources and leveraging advanced infrastructure, such as cloud-based solutions or distributed computing platforms, can alleviate resource constraints. These improvements will optimise performance, enabling the system to process large-scale datasets and deliver reliable results efficiently.

*D. Generalisability of results*

The system's generalisability stems from its use of unsupervised machine learning, allowing it to adapt to various contexts. The nature of the datasets influences results; for example, analysing cloud security datasets yields insights specific to cloud security. This adaptability ensures the system's findings remain relevant and applicable across diverse scenarios, enhancing the overall utility of the results.

*E. Toward Multi-Layered Pipeline Defense*

While this work focuses on semantic topic modelling, our long-term vision includes embedding this layer into a broader security framework. For example, CNN-LSTM models have shown strong potential for detecting temporal attack patterns in real pipeline environments [34]. Combining topic-based insight with sequence-based anomaly detection and immutable blockchain logs may enable layered trust, supporting both semantic interpretation and forensic auditability of CI/CD workflows.

## VII. CONCLUSIONS AND FUTURE WORKS

With the growing need for stronger cloud security in CI/CD pipelines, this study presents a topic modelling-based framework designed to uncover potential vulnerabilities from text-based artefacts generated throughout the software delivery lifecycle. By leveraging natural language processing, specifically Latent Dirichlet Allocation (LDA) and pLSA, the system helps extract latent security-relevant themes from logs, reports, and other pipeline documents. This contributes a novel semantic layer for early threat detection in DevSecOps workflows.

The proposed approach introduces a scalable, explainable method for understanding pipeline behaviour from an information-theoretic perspective. It serves as a foundational step toward enhancing runtime visibility, with the broader objective of surfacing misconfigurations, policy violations, or suspicious activities that traditional rule-based systems may overlook.

Future work will expand the system in several directions.

First, integrating the system with existing CI/CD tools such as Jenkins, Tekton, Travis CI would enable automated, real-time log analysis and better workflow alignment.

Second, refining pre-processing mechanisms and adding intelligent filters to eliminate semantically irrelevant but similar documents can improve accuracy and reduce false positives.

Third, expanding support for diverse file types and data formats will make the system more adaptable to different environments and sources of security-related text.

Future directions will also focus on fusing this semantic modelling layer with an AI-based anomaly detection system and blockchain-anchored log integrity. Building on prior deployments of CNN-LSTM models within Jenkins environments [28] and modular blockchain CI/CD architectures [30], this integration aims to support real-time detection, localised threat containment, and trustworthy logging in dynamic DevSecOps workflows. These directions directly address the structural and tooling gaps identified in recent CI/CD security reviews [31].

ACKNOWLEDGEMENT

We acknowledge Daniel Szabo, an undergraduate student at Western University, for his dedicated efforts in implementing the code and assisting with the technical aspects of this project. Daniel's work executing and refining these implementations was crucial to advancing the research, and their commitment to the project was influential in bringing the research to completion.